\documentclass[floatfix,superscriptaddress,a4paper,
               nofootinbib,preprint]{revtex4}

\pdfoutput=1

\usepackage{graphicx}
\usepackage{amsmath}
\usepackage{amssymb}
\usepackage{bm}

\newcommand{\hsp}{\hspace*{1pt}}
\newcommand{\hspm}{\hspace*{.5pt}}

\newcommand{\be}{\begin{equation}}
\newcommand{\ee}{\end{equation}}
\newcommand{\bel}[1]{\be\label{#1}}
\newcommand{\re}[1]{Eq.~(\ref{#1})}

\begin{document}

\title
{
Bose-Einstein condensate of alpha particles \\in the ground state of nuclear matter?
}

\author{L.~M. Satarov}
\affiliation{
Frankfurt Institute for Advanced Studies, D-60438 Frankfurt am Main, Germany}

\author{M.~I. Gorenstein}
\affiliation{
Frankfurt Institute for Advanced Studies, D-60438 Frankfurt am Main, Germany}
\affiliation{
Bogolyubov Institute for Theoretical Physics, 03680 Kiev, Ukraine}

\author{I.~N. Mishustin}
\affiliation{
Frankfurt Institute for Advanced Studies, D-60438 Frankfurt am Main, Germany}
\affiliation{
National Research Center ''Kurchatov Institute'' 123182 Moscow, Russia}

\author{H. Stoecker}
\affiliation{
Frankfurt Institute for Advanced Studies, D-60438 Frankfurt am Main, Germany}
\affiliation{
Institut f\"ur Theoretische Physik,
Goethe Universit\"at Frankfurt, D-60438 Frankfurt am Main, Germany}
\affiliation{
GSI Helmholtzzentrum f\"ur Schwerionenforschung GmbH, D-64291 Darmstadt, Germany}


\begin{abstract}
The phase diagram of isospin-symmetric chemically equilibrated mixture
of alpha particles ($\alpha$) and nucleons ($N$) is studied in the mean-field approximation.
Skyrme-like parametrization is used for the mean-field potentials as functions of
partial densities $n_\alpha$ and~$n_N$.
We find that there is a threshold value $a_*$ of the parameter $a_{N\alpha}$ which
describes the attractive interaction between \mbox{$\alpha$ particles} and nucleons. At $a_{N\alpha}<a_*$
the ground state of nuclear matter at zero temperature behaves as a pure system of interacting nucleons,
whereas at $a_{N\alpha}>a_*$ the nuclear ground state includes also a nonzero fraction of $\alpha$.
We demonstrate that the equation of state of such $\alpha$-$N$ system
includes both the first-order liquid-gas phase transitions and the Bose-Einstein condensation
of~$\alpha$~particles.
\end{abstract}

\maketitle

\section{Introduction}

As commonly accepted, see, e.g.~\cite{Wal04,Sat09}, the ground state (GS) of isospin-symmetric nuclear matter at zero temperature and pressure
is characterized by the following parameters~(\mbox{$\hbar=c=k_B=1$}):
\bel{GSp}
n_B=n_0\simeq 0.15~{\rm fm}^{-3}~,~~~~W=\frac{\varepsilon}{n_B}-m_N =W_0\simeq -15.9~{\rm MeV}\hsp,
\ee
where $n_B$ is the baryonic density, $W$ is the binding energy per baryon,
$\varepsilon$ is  {the} energy density, and $m_N$ is the nucleon mass.
The Coulomb interaction effects are assumed to be switched off.   In what follows we use
$m_N \simeq 938.9$~MeV neglecting a  small difference between the proton
and neutron masses.

The  nuclear GS is usually considered in terms of interacting nucleons, i.e., neutrons and protons.
On the other hand,
it is well known that nuclear matter has a tendency for clusterization at subsaturation densities and moderate temperatures, as observed at intermediate and high collision energies. Especially clear this was
demonstrated by nuclear fragmentation reactions which have
been extensively studied, both experimentally~\cite{Sch96,Rei10,Wad19} and
theore\-tically~\mbox{\cite{Pei91,Gro93,Bon95}}.
Particularly, $\alpha$ particles ($^4$He nuclei)  are abundantly produced in intermediate-energy heavy-ion collisions \cite{Sch17}. The $\alpha$-decay of heavy
nuclei is another indication that $\alpha$-like correlations exist in cold nuclei. The quartet-type
nucleon correlations have been introduced in~\cite{Toh01} and then studied in subsequent  publications, see, e.g.~\mbox{Refs.~\cite{Fun09,Rop14}}.
The authors claim that at moderate excitation energies such correlations may give rise to the
$\alpha$-particle condensate, analogous to the famous Hoyle state in $^{12}$C\hspm .
Below we consider a similar type of correlations, but represented  {by}
$\alpha$-particles coexisting with nucleons even at zero temperature. The focus of our present study is on the role of interaction between $\alpha$-particles and nucleons.

In recent years many theoretical models have been used to describe nuclear systems with light
clusters, see, e.g.~Refs.~\cite{Lat91,Hor06,Typ10,Hem11,Bot10,Fur17,Mis15,Pai19}.
In our previous paper~\cite{Sat19} we have  {proposed} a mean-field model with Skyrme-like interaction
potentials to describe the isospin-symmetric $\alpha$-$N$ matter under conditions of chemical equilibrium.
It was assumed that the GS of such matter contains only nucleons and no $\alpha$'s. Below we demonstrate that this model allows also another possibility, when the GS contains both nucleons and $\alpha$-particles. This is controlled by the parameter $a_{N\alpha}$ which determines the strength of attractive $\alpha N$ interaction. For $a_{N\alpha}<a_*$ where $a_*\sim 2~\textrm{GeV\hsp fm}^3$ is a certain threshold value (see below),
the GS of nuclear matter at temperature $T=0$ contains only nucleons.

In the present paper we demonstrate that for sufficiently strong $\alpha N$ attraction, \mbox{$a_{N\alpha}>a_*$},
the nuclear GS contains a nonzero fraction of $\alpha$-particles. In this case we obtain a~qualitatively different phase diagram
of $\alpha$-$N$ matter which contains both the first-order liquid-gas phase transition (LGPT) and the Bose-Einstein condensate (BEC) of $\alpha$-particles.

\section{The model}

Let us consider an iso-symmetric system (with equal numbers of protons and neutrons)
composed of nucleons and $\alpha$-particles with vacuum mass $m_\alpha\simeq 3727.3$~MeV.
In the  grand canonical ensemble the system pressure $p\,(T,\mu)$
is a function of temperature~$T$ and baryon chemical potential  $\mu$.
The latter is responsible for conservation of the baryon number.
The chemical potentials
of $N$ and $\alpha$ satisfy the relations $\mu_N=\mu,$ and $\mu_\alpha=4\mu$,
which correspond to condition of the chemical equilibrium in the $\alpha$-$N$ system
with respect to reactions~\mbox{$\alpha\leftrightarrow 4N$}.
The baryon number density~$n_B\hsp (T,\mu)=n_N+4n_\alpha$, the entropy density~$s\hsp (T,\mu)$,
and the energy density $\varepsilon\hsp (T,\mu)$ can be calculated from~$p\,(T,\mu)$ and its partial derivatives using the standard thermodynamic relations.  
Our consideration is restricted to
temperatures $T\lesssim 20~\textrm{MeV}$ to avoid complications due to production of mesons and
hadronic resonances.

In our  mean-field approach the pressure $p\,(T,\mu)$ of the $\alpha$-$N$ system  is taken in the form
\bel{tpwi}
p=p_N^{\hsp\rm id}(T,\widetilde{\mu}_N)+p_\alpha^{\hsp\rm id}(T,\widetilde{\mu}_\alpha)+\Delta p\hsp (n_N,n_\alpha)\hsp.
\ee
Here $p_i^{\rm id}$ are the ideal-gas pressures  of $i$th particles:
\bel{pid}
p^{\,\rm id}_i(T,\widetilde{\mu}_i)~=~\frac{g_i}{(2\pi)^3}\int d^{\,3}k\frac{k^2}{3E_i}\left[
\exp{\left(\frac{E_i-\widetilde{\mu}_i}{T}\right)}\pm 1\right]^{-1}~,~~~(i=N,\alpha)\hsp ,
\ee
where $E_i=\sqrt{m_i^2+k^2}$, $g_i$ is the spin-isospin degeneracy factor ($g_\alpha=1, g_N=4$)\hsp.
Upper and lower signs
in Eq.~(\ref{pid}) correspond to~$i=N$ and~$i=\alpha$, respectively.
Particle interactions in the $\alpha$-$N$ mixture are described  by introducing temperature-independent
mean-field  potentials which are parametrized in
a Skyrme-like  form. These  potentials lead to the shifts of the chemical potentials
 with respect to their ideal gas values.

 Following Ref.~\cite{Sat19}, we apply the parametrizations
\bel{tilde}
 \widetilde{\mu}_N=\mu + 2a_N n_N+2a_{N\alpha} n_\alpha -
\frac{\gamma+2}{\gamma+1}\hsp b_N \left(n_N + \xi n_\alpha\right)^{\gamma+1}\hsp ,
\ee
\bel{tilde-a}
\widetilde{\mu}_\alpha=4\mu + 2a_{N\alpha} n_N+2a_{\alpha} n_\alpha -
\frac{\gamma+2}{\gamma+1}\hsp b_N \xi \left(n_N + \xi n_\alpha\right)^{\gamma+1}\hsp ,
\ee
 and
\bel{dpit1}
\Delta p\hsp (n_N,n_\alpha)= - a_N\hsp n_N^2 - 2a_{N\alpha}n_N\hsp n_\alpha -a_\alpha\hsp n_\alpha^2
+ b_N \left(n_N+\xi  n_\alpha\right)^{\gamma+2}\hsp ,
\ee
where
$a_{N}$, $a_\alpha$, $a_{N\alpha}$, $b_N$, $\xi$, and $\gamma$ are positive model parameters.
 Using Eqs.~(\ref{tpwi})--(\ref{dpit1}) one can show that the condition of thermodynamic consistency,
$n_B=\left(\partial p/\partial\mu\right)_T$, holds for all $T$ and~$\mu$.
The BEC of $\alpha$ particles  {becomes possible} when $\widetilde{\mu}_\alpha$ reaches its maximum value~$\widetilde{\mu}_\alpha=m_\alpha$.

The terms with coefficients $a_{N}, a_\alpha, a_{N\alpha}$ in~Eqs.~(\ref{tilde})--(\ref{dpit1}) describe attractive forces,
whereas the terms proportional to $b_N$ are responsible for repulsive interactions.
Similar to Ref.~\cite{Sat19} we take the parameters $\gamma=1/6$,
\mbox{$a_\alpha= {3.83}~\textrm{GeV\hsp fm}^3$}, and $\xi= {2.01}$.
These values were fixed by  {fitting} the parameters $n_\alpha=0.036$~fm$^{-3}$ and
$W_\alpha=\varepsilon_\alpha/n_\alpha -m_N = - 12$~MeV for the~GS of~pure $\alpha$ matter~at $T=0$~\cite{Cla66,Sat17}.
The nucleon parameters $a_N= {1.17}~\textrm{GeV fm}^3$ and  $b_N= {1.48}~\textrm{GeV fm}^{7/2}$  were obtained in  Ref.~\cite{Sat19}
by fitting the GS  {characteristics} (\ref{GSp}) of the pure nucleon matter, i.e.,
assuming that it contains no $\alpha$ particles.

Thus, only one
unknown parameter is left in the parametrization~(\ref{tilde})--(\ref{dpit1}), namely the cross-term
coefficient $a_{N\alpha}$ which determines the $\alpha N$ attraction
strength. It will be demonstrated below that this parameter plays a crucial role in thermodynamics
of $\alpha$-$N$ matter.  {In Ref.~\cite{Sat19} we have shown} that the results are qualitatively different for $a_{N\alpha}$  smaller  {or} larger than the threshold value
\bel{thrc1}
a_*=-\frac{2}{n_0}\hsp (W_0+B_\alpha)+
\frac{\footnotesize 1}{\footnotesize 2}\hsp\frac{\footnotesize\gamma+2}{\footnotesize\gamma+1}\hsp b_N\hsp\xi\hsp n_0^\gamma\simeq 2.12~\textrm{GeV\hsp fm}^3\hsp ,
\ee
where $B_\alpha=m_N-m_\alpha/4\simeq 7.1~\textrm{MeV}$ is the binding energy per baryon of the $\alpha$ nucleus\hspm\footnote
{
Note that $a_{N\alpha}=a_*$ satisfies approximately the 'mixing rule'
$a_{N\alpha}\simeq\sqrt{a_N\cdot a_\alpha}\hsp$,
found experimentally~\cite{Pat82} for attractive mean-field interactions in
binary mixtures of molecular liquids.
}.
\begin{figure}[htb!]
\centering
\vspace*{-5mm}\includegraphics[trim=3cm 8cm 3cm 8.7cm,height=0.4\textheight]{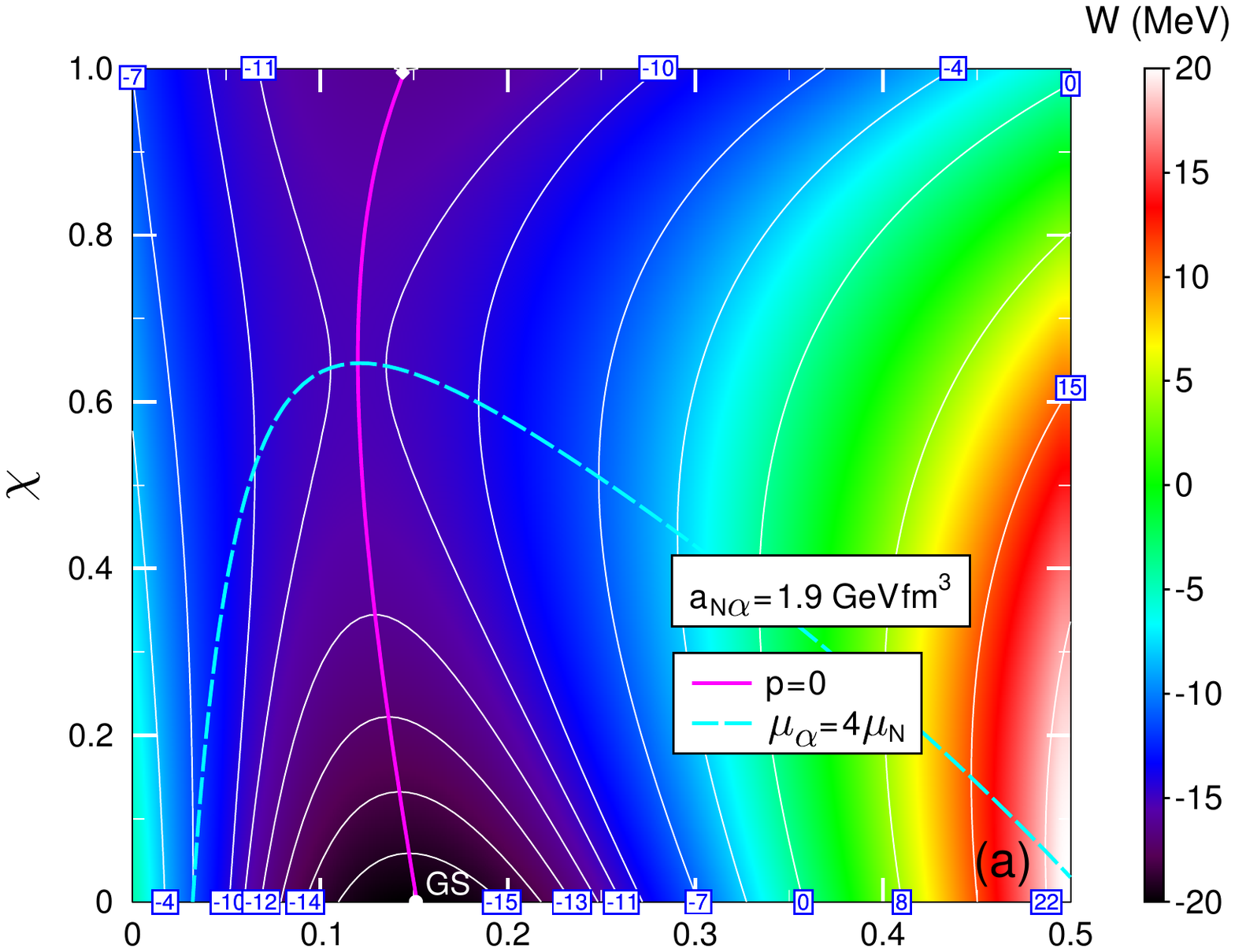}
\includegraphics[trim=3cm 8cm 3cm 8.7cm,height=0.4\textheight]{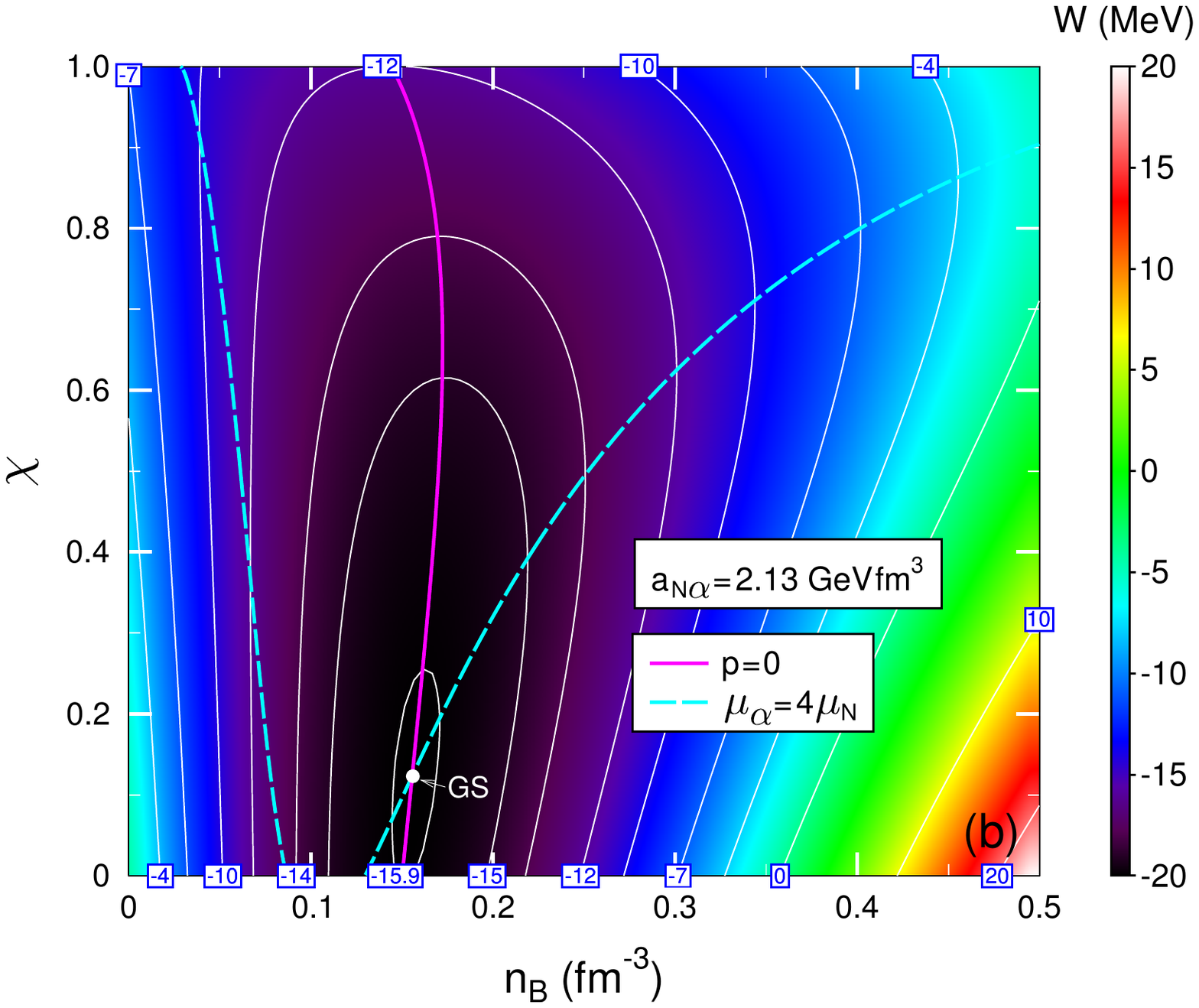}
\caption{
The binding energy per baryon $W$ for cold  $\alpha$-$N$ matter on the
$(n_B,\chi)$ plane.
White contours correspond to constant $W$ values (given in MeV inside white boxes). Panel~(a) shows the results for $a_{N\alpha}=1.9~\textrm{GeV\hsp fm}^3<a_*$.
The GS corresponds to a pure nuc\-leon matter (\mbox{$n_\alpha=0$})  {with parame\-ters given
in}~\re{GSp}.  {The} metastable state of a pure $\alpha$-matter is shown by the white diamond at the~\mbox{$\chi=1$} axis. The results in (b) are obtained for
\mbox{$a_{N\alpha}=2.13~\textrm{GeV\hsp fm}^3 > a_*$}.  In this case the system has only one minimum of $W$ and the GS contains both nucleons and~$\alpha$'s.
}\label{fig-T0}
\end{figure}

 {To constrain the parameter $a_{N\alpha}$, we have compared our results with those obtained within the virial expansion approach~\cite{Hor06} for the iso-symmetric $\alpha$-$N$ matter. The latter is justified
in the domain of small densities $n_N$ and $n_\alpha$ where one can calculate thermodynamic properties of nuclear matter by using empirical phase shifts of $NN, N\alpha$\hspm, and $\alpha\alpha$~scattering. In this analysis, $a_{N\alpha}$ has been varied in the interval}\hspm\footnote
{
 {A similar analysis has been made in Ref.~\cite{Sat19} for $a_{N\alpha}=1$ and
$1.9~\textrm{GeV\hsp fm}^3$.}
}
 {from~1 to~2.5~$\textrm{GeV\hsp fm}^3$. The best agreement with the virial approach at~$T\sim 2~\textrm{MeV}$ has been achieved for $a_{N\alpha}$ close to $a_*$, but with a significant uncertainty of about~$\pm 10\% $.}
 {From this analysis we can not conclude which option, $a_{N\alpha}>a_*$ or $a_{N\alpha}<a_*$, is more realis\-tic and, therefore, consider both of them.}

\section{Results for $T=0$}

At $a_{N\alpha}<a_*$
the~GS  {of} $\alpha$-$N$ mixture corresponds to  {a pure nucleonic} matter ($n_\alpha=0$)
which satisfies the GS properties (\ref{GSp}). This  {option} has been studied earlier in Ref.~\cite{Sat19}. The  {results for} $a_{N\alpha}=1.9~\textrm{GeV\hsp fm}^3$  {are presented} in Fig.~\ref{fig-T0}(a)
where contours of $W$ are shown in the $(n_B,\chi)$ plane.  {Here} $\chi \equiv 4n_\alpha/n_B$ is  {the} fraction of  {nucleons} carried
by~$\alpha$~particles. The red solid curve is the line of zero pressure $p=0$.
Besides the GS (\ref{GSp}), there is another state with a local minimum of
the energy per baryon at~$\chi=1$ and $n_N=0$, which corresponds
to the~pure $\alpha$ matter, considered earlier in Ref.~\cite{Cla66}. This state is metastable because it has a~smaller binding energy  $|W|$ as compared to the~pure nucleonic matter.  {In the~($n_B,\chi$) plane the} two minima are separated  by a potential barrier (see the dashed line in Fig.~\ref{fig-T0}(a)).

\begin{figure}[htb!]
\centering
\includegraphics[trim=2cm 8cm 3cm 9cm,width=0.48\textwidth]{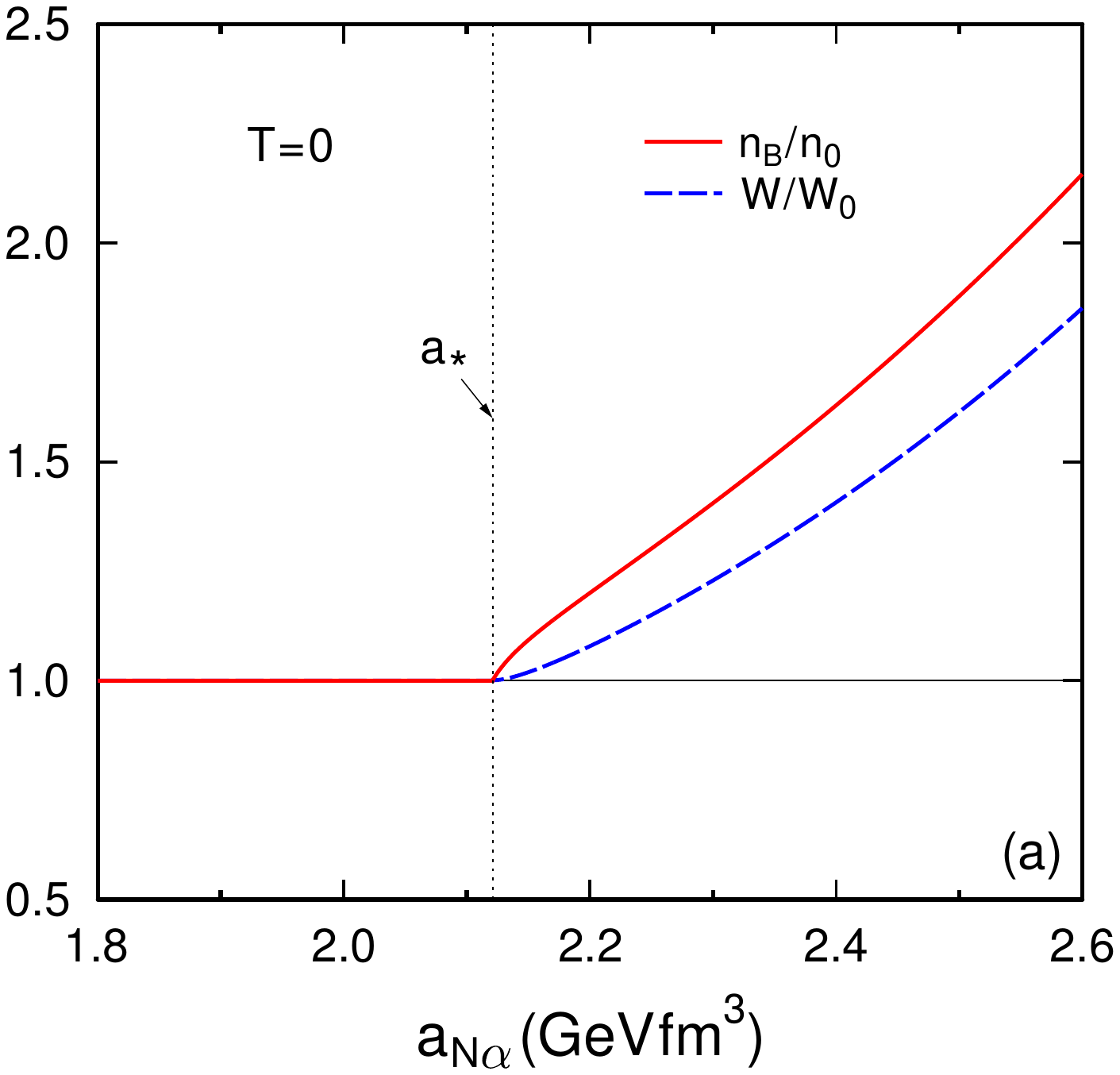}
\includegraphics[trim=2cm 8cm 3cm 9cm,width=0.48\textwidth]{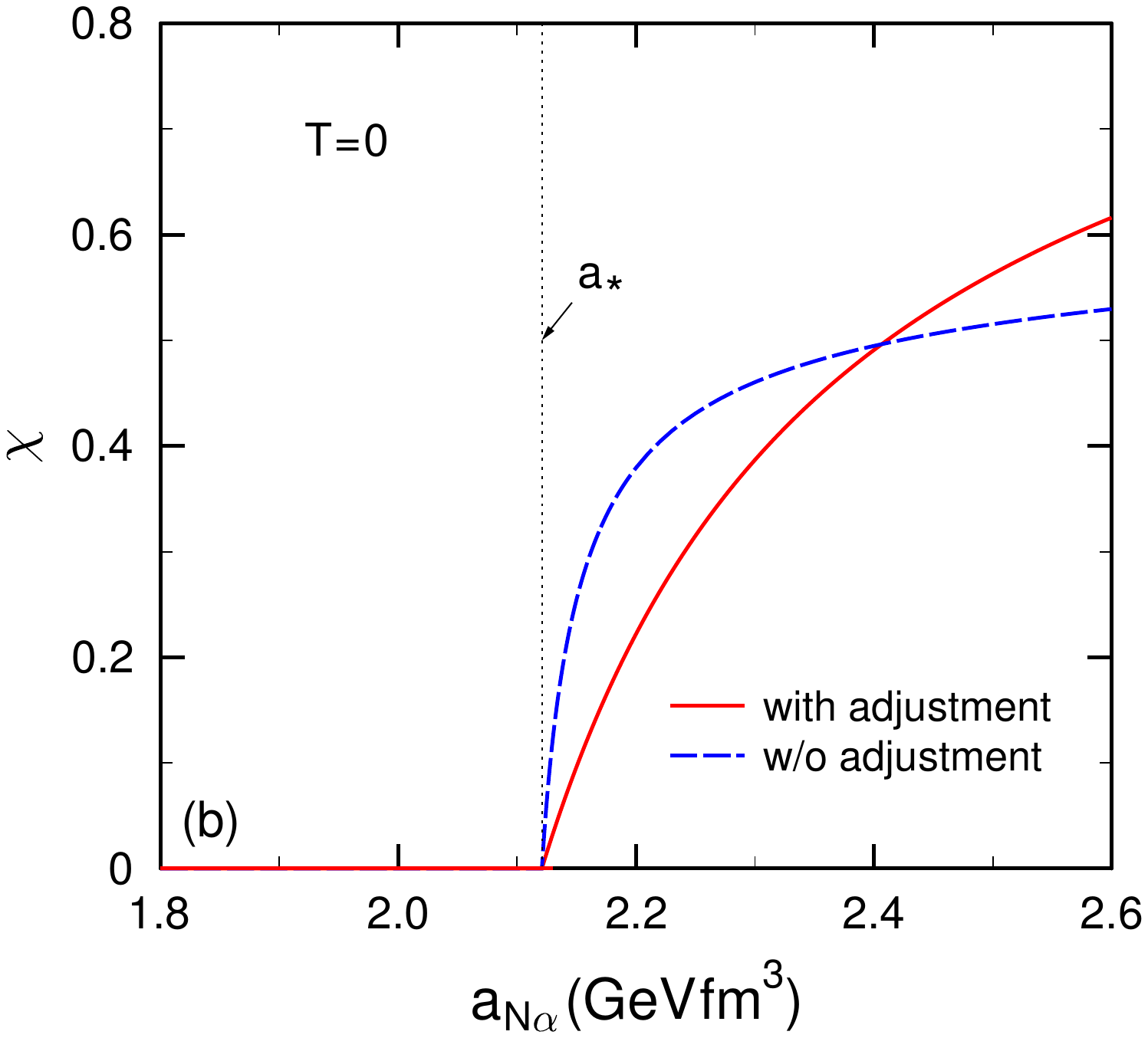}
\caption{(a) Relative change of baryon density (the solid line) and binding energy (the dashed curve)
in the GS of $\alpha$-$N$ matter as a function of $a_{N\alpha}$.
\mbox{(b) Fraction} of $\alpha$'s in the GS of $\alpha$-$N$ matter at different~$a_{N\alpha}$.
The solid (dashed) line is obtained with (without) readjusting $a_N,b_N$\hsp .
}\label{fig-W}
\end{figure}

 {In the present paper we consider a} new possibility  {which} arises at  \mbox{$a_{N\alpha}>a_*$}.
In this case, the energy per baryon has only one \mbox{minimum} in the $(n_B,\chi)$ plane
at $\chi >0$,  and  {the} GS contains both the Fermi distribution of nucleons as well as the Bose condensate of $\alpha$-particles.
An example of such a~system is shown in Fig.~\ref{fig-T0}(b) for \mbox{$a_{N\alpha}=2.13~\textrm{GeV\hsp fm}^3$}.
Indeed, one can see that now the GS minimum of $W$ is shifted from the $\chi=0$ axis to $\chi\simeq 0.12$\hspm.  {In the considered case both pure nucleonic- and pure
$\alpha$-matter appear as unstable states.}

According to our calculation, at $a_{N\alpha} >a_*$ the GS parameters $n_B$ and $|W|$ increase monotonously with $a_{N\alpha}$ as shown in Fig.~\ref{fig-W}(a).
Therefore,  {in this case} the GS of the $\alpha$-$N$ mixture does not satisfy the conditions (\ref{GSp}) since  $n_B>n_0$ and $|W|>|W_0|$\hspm.  {To}
fulfill the empirical constraints (\ref{GSp})  {we} readjust the nucleon interaction parameters $a_N$ and $b_N$\hspm\footnote
{
We  find that readjusted parameters  {$a_N$ and $b_N$} increase almost linearly with $a_{N\alpha}$. The {increase factors are about 1.25 and 1.4} when $a_{N\alpha}$ changes from
$a_*$ to  {$1.1\hsp a_*$}\hspm.
}.
The results of such calculation are shown in Fig.~\ref{fig-W}(b).
For example, at $a_{N\alpha}= 2.13$~GeV~fm$^3$ the readjusted coefficients $a_N$ and $b_N$ are increased by
about 1\% as compared to their values at $a_{N\alpha}<a_*$,  {but} the
fraction of alphas, $\chi$, dropped significantly, from~0.12 to 0.04.

\section{Phase diagram of $\alpha$-$N$ matter}

\begin{figure}[htb!]
\centering
\includegraphics[trim=2cm 8cm 3cm 9cm,width=0.48\textwidth]{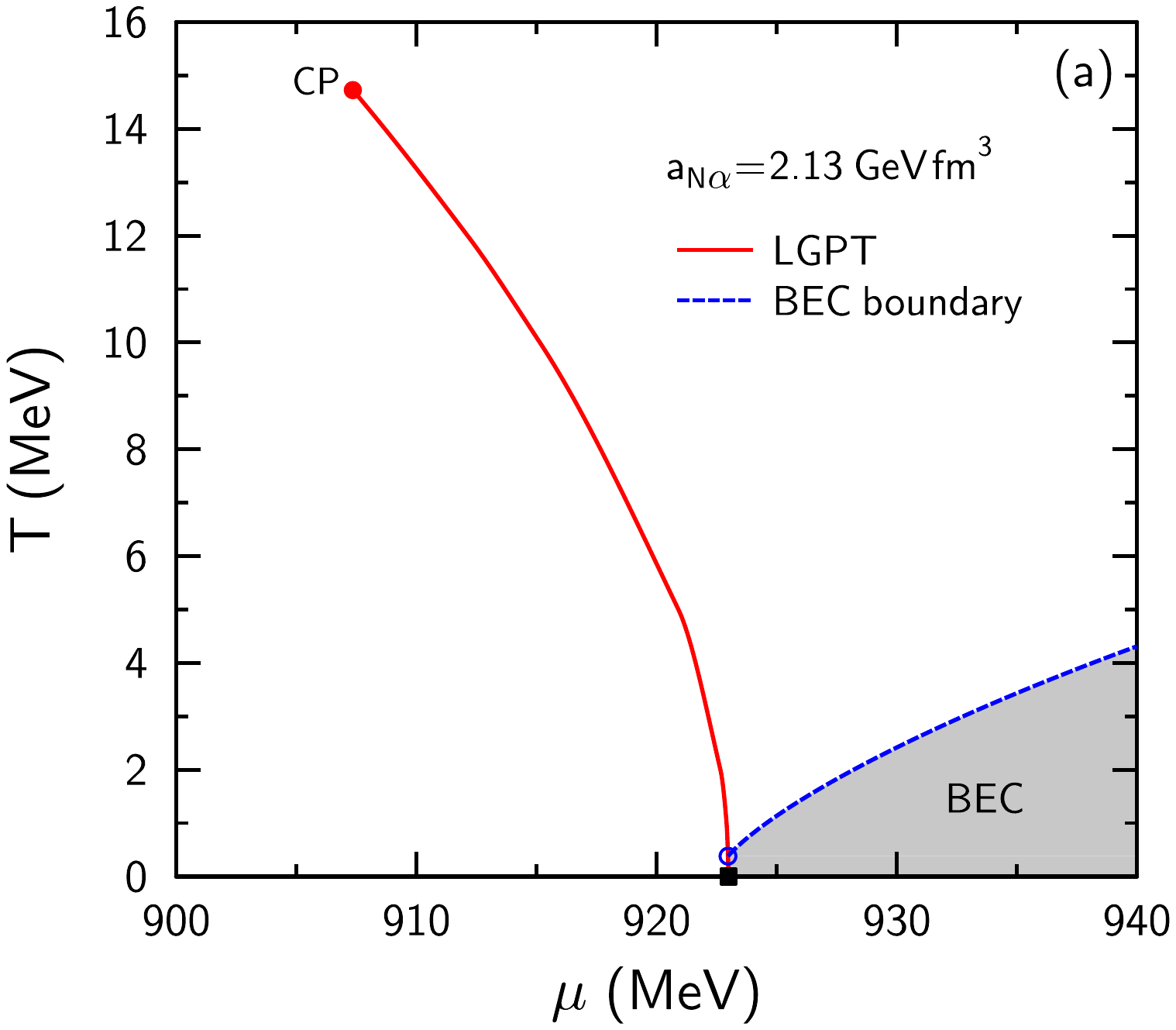}
\includegraphics[trim=2cm 8cm 3cm 9cm,width=0.48\textwidth]{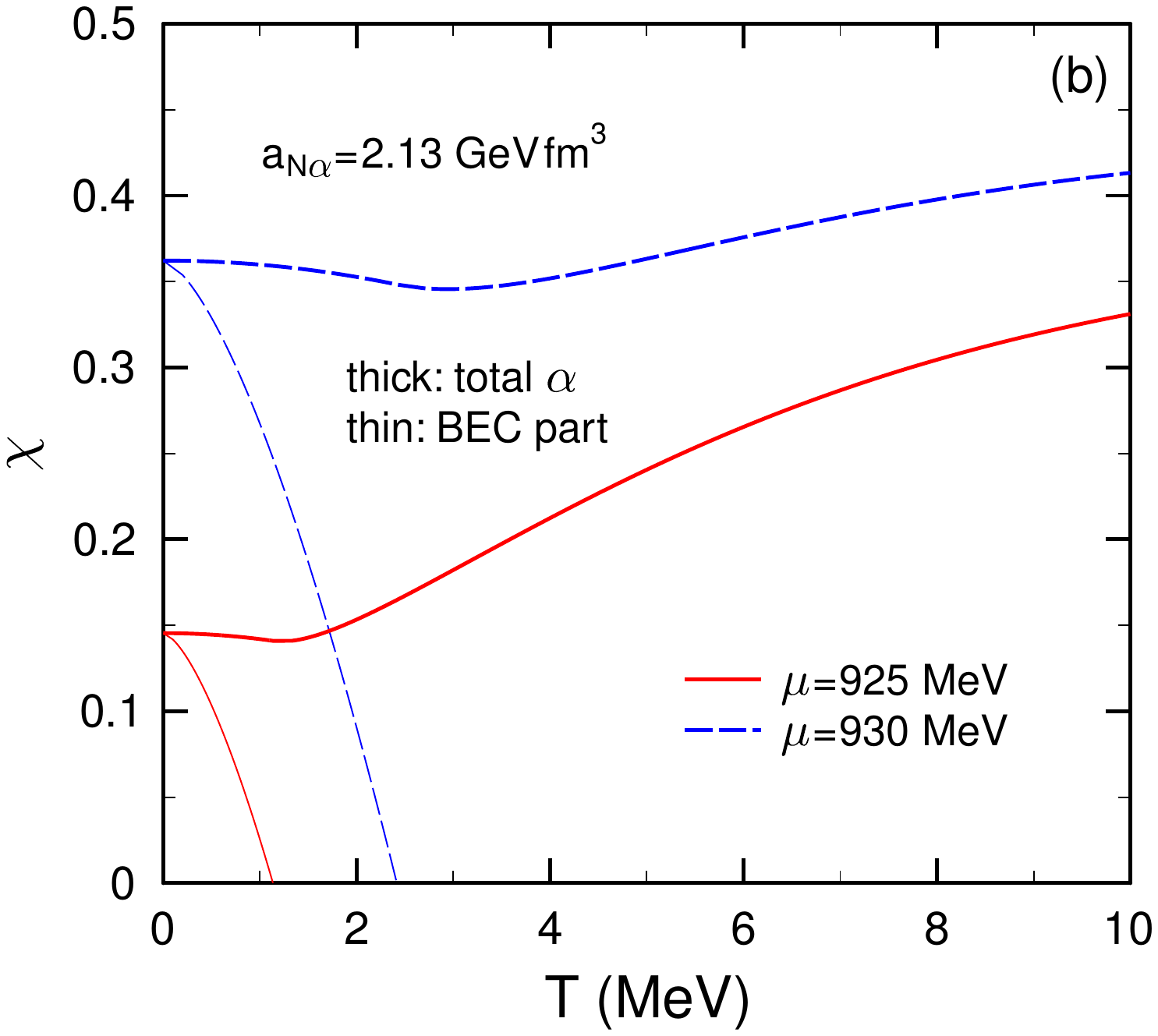}
\caption{ {(a):}~Phase diagram of $\alpha$-$N$ matter for $a_{N\alpha}= 2.13$~GeV~fm$^3$
on the $(\mu,T)$ plane. The full square shows the GS position, whereas the full and open circles corresponds to the critical and triple points, respectively.  Shading marks states with BEC
of $\alpha$ particles.
(b): Fractions of $\alpha$-particles in $\alpha$-$N$ matter as functions of $T$ for $a_{N\alpha}= 2.13~\textrm{GeV\hsp fm}^3$.
The  {solid and dashed} lines correspond to  $\mu=225$ and 230~MeV, respectively.
The  {thick} lines show fractions of all $\alpha$'s, while the  {thin} ones give the  {contribution} of the condensate.
}\label{PTD}
\end{figure}

\begin{figure}[htb!]
\centering
\includegraphics[trim=2cm 7.5cm 3cm 8.5cm,width=0.7\textwidth]{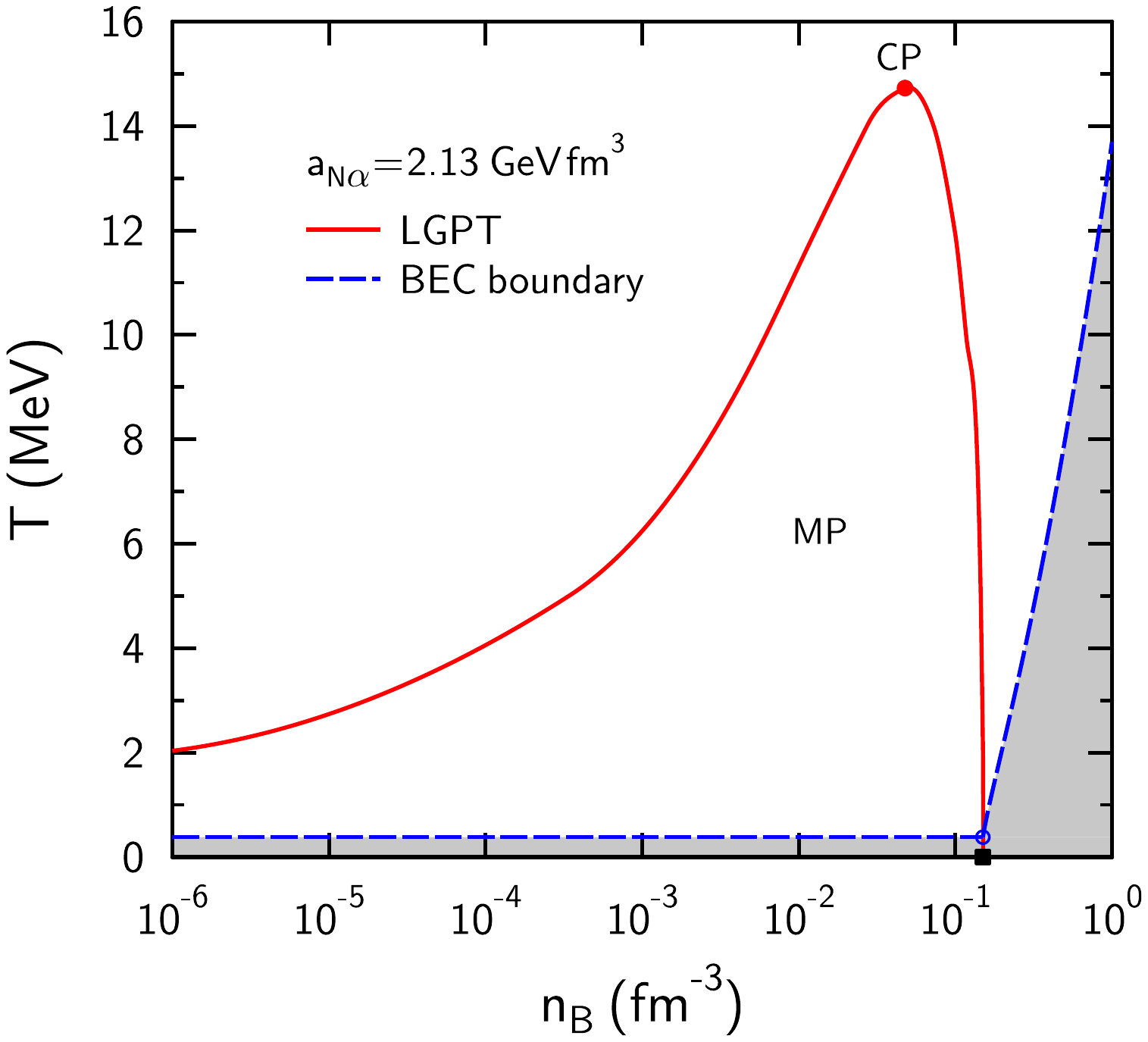}
\caption{Same as Fig.~\ref{PTD}(a) but for the $(n_B,T)$ plane\hsp. MP denotes the
mixed-phase region.
}\label{Fig4}
\end{figure}

In~Figs.~\ref{PTD}(a)~and ~{\ref{Fig4}} we present the
phase diagram of $\alpha$-$N$ matter for \mbox{$a_{N\alpha}= 2.13$~GeV~fm$^3 >a_*$}
on the $(\mu,T)$ and $(n_B,T)$ planes, respectively.  {Note that the squares show positions of the GS in both diagrams.}
The model predicts
the first-order phase transition of the liquid-gas type with following parameters of critical point (CP)
\bel{crpar}
T_{\rm CP}\simeq ~14.7~\textrm{MeV},~~~\mu_{\rm CP}\simeq ~907~\textrm{MeV},~~~n_{\rm BCP}\simeq 0.048~\textrm{fm}^{-3}, ~~~\chi_{\rm CP}\simeq 0.19\hsp.
\ee
In our calculations we use the Gibbs conditions of the phase equilibrium~\cite{Lan75}. We also predict
the BEC phase of alpha particle. The regions of phase diagrams containing states with $\alpha$ condensate are shown in Figs.~\ref{PTD}(a) and \ref{Fig4} by shading.

In the present scenario, the BEC states are thermodynamically stable. This is different from the case considered in our previous work~\cite{Sat19} where the BEC states of $\alpha$'s
appeared as a metastable phase. One can see that these states are located
to the right-hand side from the phase transition line started from the triple point in the $(\mu,T)$ plane.
We find rather low temperature of the triple point, $T_{\rm TP}\simeq  {0.38}~\textrm{MeV}$.
Qualitatively similar results are found for other values of $a_{N\alpha}>a_*$.

As one can see from Fig.~\ref{Fig4} the BEC states exist even in the two-phase coexisting region
at~$n_B<n_0$\hspm . Here the system splits into domains of higher (liquid) and smaller (gas) densities. In this case the $\alpha$ condensate is localized in liquid domains (droplets) which have baryon density close to $n_0$ irrespective of the baryon density $n_B$ in the MP. On the other hand, the condensate does not appear in the pure gas phase located on the left from the~MP region.

It is interesting to estimate the fraction of $\alpha$-particles in the BEC phase
as a function of temperature. According to our present model, at $T\to 0$ and $\mu>m_N+W_0\simeq 923~\textrm{MeV}$ all~$\alpha$'s are in the condensate, but at nonzero temperature they partly go to the non-condensed phase.
A~more detailed information is given in~Fig.~ {\ref{PTD}(b)} where we
show the temperature dependence of~$\chi$ for two fixed values of $\mu$\hspm.
The thick lines represent the total fraction of $\alpha$'s and the thin ones give the fraction of $\alpha$'s in the BEC phase. At considered values of $\mu$ this fraction decreases with~$T$ and vanishes at the BEC boundary shown by the dashed line in~Fig.~\ref{PTD}(a).

\begin{figure}[htb!]
\centering
\includegraphics[trim=2cm 8cm 3cm 9cm,width=0.48\textwidth]{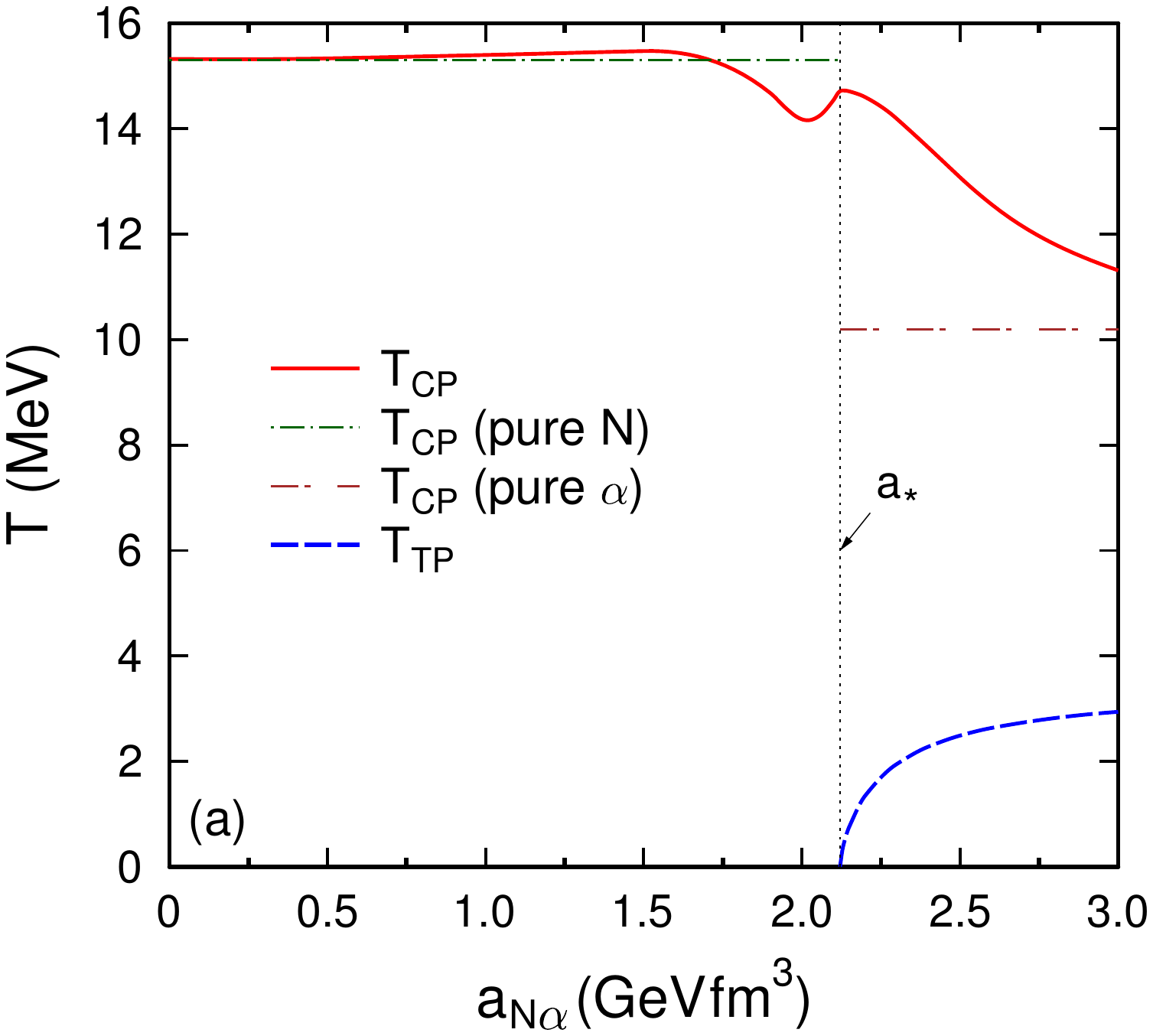}
\includegraphics[trim=2cm 8cm 3cm 9cm,width=0.48\textwidth]{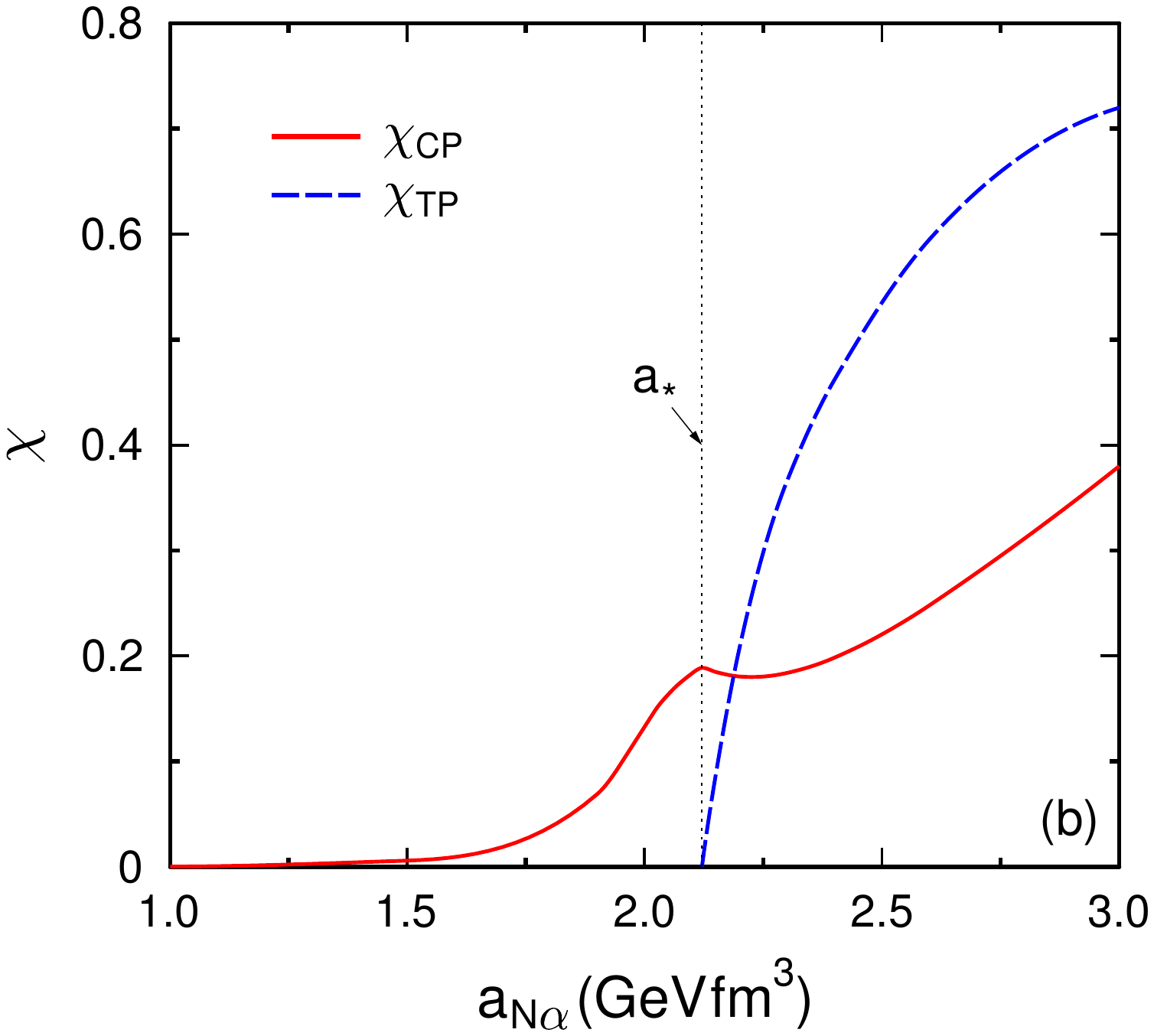}
\caption{ {(a): The temperatures at the CP (the solid line) and TP (the dashed curve)
for $\alpha$-$N$ matter as functions of~$a_{N\alpha}$.
The horizontal upper and lower dash-dotted lines correspond to the critical temperatures of the pure
nucleonic- and $\alpha$-matter, respectively. (b): Same as (a), but for relative fractions
of $\alpha$'s\hspm.}
}\label{Fig5}
\end{figure}
 {To study sensitivity of the results to the coefficient of $\alpha N$ attraction, we calculated characteristics of the CP and TP at different $a_{N\alpha}$}\hspm\footnote
{
 {Some results for~\mbox{$a_{N\alpha}<a_*$}
have been presented earlier in Ref.~\cite{Sat19}.}
}.
 {As one can see in Fig.~\ref{Fig5}(a),  the critical temperatures $T_{\rm CP}$ at small and large $a_{N\alpha}$ are close to those for a pure
nuc\-leonic~(\mbox{$T_{\rm CP}\simeq 15.3~\textrm{MeV}$}) and pure $\alpha$ ($T_{\rm CP}\simeq 10.2~\textrm{MeV}$)
matter, respectively\hspm. The values of $T_{\rm CP}$ and $\chi_{\rm CP}$ show a
non-monotonous behavior at $a_{N\alpha}\sim a_*$. Both these quantities have local maxima at~$a_{N\alpha}=a_*$. On the other hand, $T_{\rm TP}$ and $\chi_{\rm TP}$ are increasing functions of $a_{N\alpha}$. One can see that $\chi_{\rm TP}$ rapidly raises with~$a_{N\alpha}$ above the threshold $a_{N\alpha}=a_*$.}

We  would like to add a comment  {on} Ref.~\cite{Zha19} where the authors consider the
$\alpha$~condensation at low temperatures ($T\lesssim 1~\textrm{MeV}$) and densities ($n_B\lesssim 0.1\hsp n_0$), but assuming that the $\alpha$-$N$
system is homogeneous. However, our calculation has shown, that under such conditions the nuclear matter {exists} in a highly inhomogeneous liquid-gas phase.
As seen in our Fig.~\ref{Fig4}, a homogeneous gas-like phase at $T\lesssim 1~\textrm{MeV}$ appears only at extremely small densities $n_B\ll n_0$. According to our model, the BEC does not occur in this low density domain.

\section{Conclusions}

 {We} have considered  {a} possibility that the Bose-Einstein condensate
of $\alpha$-particles may coexist with nucleons in the ground state of cold iso-symmetric nuclear matter.
In our  {Skyrme-like} mean-field model this possibility arises when attractive \mbox{$\alpha N$~interaction} is strong enough. We have investigated the phase diagram  {of $\alpha$-$N$ matter}~at finite temperatures in
a~broad interval of baryon densities. It turns out that such a system has  {both} the liquid-gas phase transition  {and} the BEC of $\alpha$-particles.  {It is interesting that} the BEC phase appears also in the liquid-gas coexistence region  {at temperatures below the triple point}. This {picture} differs significantly from that considered in
Ref.~\cite{Sat19}, where smaller $\alpha$-$N$ couplings have been  {used} and the $\alpha$ condensate appeared  {only} as a~metastable phase.

In this paper we  {have} considered an idealized system consisting
of nucleons and \mbox{$\alpha$-particles}. Of course, other light clusters and  {heavier} fragments
can play an important role in  {intermediate-energy} heavy-ion collisions and in
astrophysical processes,  {such as super\-nova explosions and neutron-star merges. We plan to include such
clusters in future calculations.}

\begin{acknowledgments}
We are thankful to D.~V. Anchishkin, A.~Motornenko, and V.~Vovchenko for fruitful discussions.
L.M.S.~acknowledges the financial support from the Frankfurt Institute for Advanced Studies.
 The work of M.I.G. is supported by the Goal-Oriented Program of Cooperation
between CERN and National Academy of Science of Ukraine 'Nuclear Matter under Extreme
Conditions' (agreement CC/1-2019, No.~0118U005343).
I.N.M. thanks for the support of the Helmholtz International Center for  {FAIR}.
H.St.~appreciates the support from J.~M.~Eisenberg Laureatus chair.
\end{acknowledgments}

\end{document}